\documentstyle[12pt,a4wide]{article}

\newcommand{\be}{\begin{equation}}
\newcommand{\ee}{\end{equation}}
\newcommand{\bea}{\begin{eqnarray}}
\newcommand{\eea}{\end{eqnarray}}
\newcommand{\ba}{\begin{array}}
\newcommand{\ea}{\end{array}}
\newcommand{\nn}{\nonumber}

\newcommand{\wt}[1]{\widetilde{{#1}}}
\newcommand{\wh}[1]{\widehat{{#1}}}
\newcommand{\quark}{\langle\,\bar q q \,\rangle}

\newcommand{\gemischt}{\langle\,\bar q \sigma g G q \,\rangle}
\newcommand{\gev}{\mbox{~GeV}}

\newcommand{\ga}{\gamma}
\newcommand{\gam}{\Gamma}
\newcommand{\dm}{\Delta m}

\newcommand{\ds}{\Delta s}
\newcommand{\al}{\alpha}

\newcommand{\bef}{\begin{figure}}
\newcommand{\enf}{\end{figure}}
\newcommand{\intl}{\int\limits}

\def\eq.(#1){eq.~(\ref{#1})}
\def\im{{\rm Im}\,}
\def\w{\omega}

\def\go{\widetilde{\Gamma}{}^{(0)}}
\def\gi{\widetilde{\Gamma}^{(1)}}
\def\ep{\epsilon}
\def\vl{v_+}
\def\vt{v_-}

\def\za{p}
\def\zb{q}
\def\N{{\cal J}}

\def\e{{\rm e}{}}
\def\medskip{\vskip 0.7cm}
\def\ALPI{\left({\alpha_s\over\pi}\right)}
\def\Z{\zeta}

\begin{document}
\begin{titlepage}
\renewcommand{\thefootnote}{\fnsymbol{footnote}}
\begin{flushright}
HD--THEP--92--40
\end{flushright}
\vskip0.4cm
\begin{center}
\boldmath
{\large\bf THE ISGUR-WISE FUNCTION TO $O(\alpha_s)$ FROM SUM\\[10pt]
RULES IN THE HEAVY QUARK EFFECTIVE THEORY}
\unboldmath
\vskip1cm
E. Bagan\footnote{A. v.\ Humboldt fellow. On leave of absence from
Grup de F\'\i sica Te\`orica,
Departament de F\'\i sica,
Universitat Aut\`onoma de Barcelona,
E-08193 Bellaterra (Barcelona), Spain, until 30{\em th.\ } Sep.\ 92},
Patricia Ball\footnote{
Adress after September 1992: Physik Department, TU M\"{u}nchen,
D-8046 Garching bei M\"{u}nchen, FRG}\\
\vskip1cm
Institut f\"ur Theoretische Physik\\
Universit\"at Heidelberg\\
Philosophenweg 16\\
D-6900 Heidelberg, FRG\\
\vskip1cm
P. Gosdzinsky\\[1cm]
Grup de F\'\i sica Te\`orica\\
Departament de F\'\i sica\\
Universitat Aut\`onoma de Barcelona\\
E-08193 Bellaterra (Barcelona), Spain\\[1cm]
{\bf Abstract:\\[5pt]}
\parbox[t]{\textwidth}{
Radiative corrections to both perturbative and non-perturbative
contributions are added to existing calculations of the
Isgur-Wise function $\xi_{IW}$. To this end, we develop a method for
calculating
two-loop integrals in the heavy quark effective theory involving two
different scales. The inclusion of $O(\alpha_s)$ terms causes
$\xi_{IW}$ to decrease as compared to the lowest order result and
shows the importance of quantum effects. The slope parameter $\rho^2$
violates the bound given by de Rafael and Taron.}\\
\vspace{1cm}
{\em Submitted to Phys. Lett. B.}
\end{center}
\renewcommand{\thefootnote}{\arabic{footnote}}
\setcounter{footnote}{0}
\end{titlepage}
\newpage

\section{Introduction}

The heavy quark effective theory (HQET) \cite{georgi,wisger} now seems
established as a useful tool for investigating the low energy regime of
QCD in the limit of infinitely heavy quarks. New flavour symmetries
appearing in the Lagrangian allow for a considerable simplification of
the treatment of transition amplitudes of heavy particles. Since the
heavy quark decouples from the light degrees of freedom, the (heavy)
meson to (heavy) meson amplitudes, for instance, are completely
determined by a single universal function, the Isgur-Wise function
$\xi_{IW}$ \cite{wisger}. $\xi_{IW}$ describes the dynamics of the
light quark cloud and depends on the ``momentum-transfer'' $y:=
v\cdot v'$, where $v$ and $v'$ are the four-velocities of the two
heavy mesons, respectively. It satisfies the normalization condition
$\left. \xi_{IW}(y)\right|_{y=1} = 1$ and one can show that the slope
at the normalization point $y=1$ is determined to be $\rho^2 := {}-\left.
d\xi_{IW}(y)/dy \right|_{y=1} \geq 1/4$ \cite{bjorken}. In addition,
it has been claimed that $\rho^2 \leq 0.45$ \cite{raftaron}. As far as
 we know, the latter bound is in contradiction to almost all
parametrizations to be found in the literature \cite{neubert_vcb,ball}.
In view of this discrepancy it seems worthwile to address the problem
by means of HQET sum rules which were first developed for the (heavy)
meson-to-vacuum transition amplitude
\cite{broadgro2,B3D,neubert_physrev} and yield nice agreement with
lattice calculations. Since the former calculations stress the
importance of radiative corrections, it is of great interest to include
them in the Isgur-Wise function as well, whereas they have been omitted
from existing sum rule calculations of $\xi_{IW}$
\cite{ball,neubert_physrev,rady}. In this letter we now present a
calculation of the Isgur-Wise function in the framework of HQET sum
rules including radiative corrections up to $O(\alpha_s)$.

Our paper is organized as follows: in Sect.~2 we develop a method for
calculating two-loop integrals in the HQET involving two different
scales and velocities as a series in $r:= (y-1)/(y+1)$ and give our
result to $O(\al_s)$ for the three-point function relevant to the
determination of $\xi_{IW}$. In Sect.~\ref{evaluation} we derive a sum
rule for the renormalization group invariant quantity
$\wh{\xi}_{IW}$. Finally, Sect.~\ref{discussion} is devoted to the
discussion of our results.

\section{Two-loop integrals in the HQET}\label{2-loop ints}

In order to obtain the Isgur-Wise function within the framework of HQET
 sum rules we first need to evaluate the three-point function
\be\label{eq:correlationF}
\wt\Gamma (\w=\wt{q}\cdot v ,\w'=\wt{q}'\cdot v',y)\:(v+v')_\mu
=
i^2\!\!\int\!\! d^4x\, d^4y\,
e^{i\tilde{q}\cdot x-i\tilde{q}'\cdot y}\,
\langle\,0\,|\,T\wt{J}_5(x) \wt{V}_\mu(0) \wt{J}_5^\dagger(y)\,|
\,0\,\rangle\, ,
\ee
with the HQET-currents $\wt{J}_5 = \bar q i\ga_5 h_v$ and $\wt{V}_\mu
=\bar{h}_v \ga_\mu h_{v'}$, where $h_v$ denotes a heavy quark field in
the HQET with four-velocity $v$ and $q$ denotes a light quark field.

In this section we shall be mainly concerned with the
${O}(\al_s)$ (two-loop) correction, $\wt{\Gamma}{}^{(1)}_{\rm pert}$,
to the perturbative part of the three-point function. Throughout this
section we will drop the subscript {\em pert}, for no confusion can
arise. After some algebra, one can write $\wt{\Gamma}{}^{(1)}$ as a
combination of two-loop integrals of the form
\be
\int\!\! d^D\!l\, d^D\!k  \left({1\over l^2}\right)^a\!\!
\left({1\over k^2}\right)^b  \left({1\over (l-k)^2 }\right)^c \!\!
\left({1\over v\cdot l+\omega}\right)^n \!\!
\left({1\over v'\cdot l+\omega'}\right)^{n'}\!\!
\left({1\over v\cdot k+\omega }\right)^m \!\!
\left({1\over v'\cdot k+\omega'}\right)^{m'},
\label{2loop Int}
\ee
where $D=4-2\epsilon$ is the space-time dimension.
We have not succeeded in writing the result of these integrations in a
closed form
convenient enough for numerical analysis.
We have
overcome this difficulty by expanding $v'$ and $v$ around $(v'+v)/2$,
thus obtaining
an expansion of the integrals in powers of~$r=(y-1)/(y+1)$.
For the numerical analysis of Sect.~\ref{discussion}, it will be
sufficient
to truncate the series at
order $r^2$. Since this expansion technique
can certainly be 
useful in other ${O}(\alpha_s)$ calculations, we shall present it here
briefly.

We begin by introducing the variables
$
\vl=(v+v')/2
$ and
$
\vt=(v-v')/ 2
$.
 Because of $v^2={v'}^2=1$, we have
$
\vl\cdot \vt=0
$.
We now expand
all
$v$- and $v'$-dependent factors
in \eq.(2loop Int)
in powers of $\vt\cdot k$ or $\vt\cdot l$.
To get the required accuracy, we need to keep only the
first five terms of the expansion.
We end up with
integrals of the form
\begin{equation}
\N=\int{d^Dl\; d^Dk\,(\vt\cdot l)^{\za} (\vt\cdot k)^{\zb} \over
(l^2)^a (k^2)^b
((l-k)^2)^c (\vl\cdot l+\omega)^n (\vl\cdot l+\omega')^{n'}
(\vl\cdot k+\omega)^m (\vl\cdot k+\omega')^{m'}}.
\label{star}
\end{equation}
This expansion removes the main source of trouble, namely the presence
of two different velocities in the denominators.
We next scale the loop variables as
\begin{equation}
l^{\mu}\rightarrow l^{\mu}\,\sqrt{ \vl^2 }~,\qquad
k^{\mu}\rightarrow k^{\mu}\,\sqrt{ \vl^2 },
\label{scaling l k}
\end{equation}
and introduce the notation $\hat v^\mu\equiv\vl^\mu/\sqrt{ \vl^2 } $,
so that $\hat v^2=1$. Eq.~(\ref{star}) now reads
\begin{eqnarray}
{\cal J}\!\!
&=&
\!\!
\left(   {1+y\over 2}   \right)^{(2a+2b+2c-2D-\za-\zb)/2}
\vt^{\mu_1}\cdots \vt^{\mu_{\za}} \vt^{\nu_1}\cdots  \vt^{\nu_{\zb}},
\label{star index}
\\
& \times &
\!\!\int
{    d^Dl\; d^Dk\;\;
    l_{(\mu_1}\cdots l_{\mu_{\za}} k_{\nu_1}\cdots  k_{\nu_{\zb})}
 \over
    (l^2)^a (k^2)^b
    ((l-k)^2)^c (\hat v\cdot l+\omega)^n (\hat v\cdot l+\omega')^{n'}
(\hat v\cdot k+\omega)^m (\hat v\cdot k+\omega')^{m'},
}
\nn
\end{eqnarray}
where the indices in parenthesis are to be symmetrized.
The integral in the last equation must be
of the form
$\sum_k{\cal A}_k({\cal T}_k)_{\mu_1\cdots\nu_q}$
where
$({\cal T}_k)_{\mu_1\cdots\nu_q}$
are totally symmetric tensors built up from $g_{\mu \nu}$ and
$\hat v_\alpha$. Note that those tensors containing
$\hat v_{\mu}$, $\hat v_{\nu}$
can be ignored, since $\hat v\cdot \vt$=0.
As a consequence, ${\cal J}\!\!_{(\za+\zb=odd)}=0$, whereas one can
show that ${\cal J}\!\!_{(\za+\zb=2n)} =O( r^n)$. Thus, only
the cases $\za+\zb=0$, $\za+\zb=2$ and $\za+\zb=4$ need be
considered. The coefficients ${\cal A}_k$ can be
obtained by the usual procedure of contracting all indices of the
integral in~\eq.(star index)
with the appropriate number of metric tensors, $g^{\mu \nu}$, and
velocities, $\hat v^\alpha$.
This will provide a set of equations for ${\cal A}_k$ which can be
solved in terms of
scalar integrals of the form
\begin{equation}
{I^{(')}}(a,b,c,p,q)=\int{d^Dl d^Dk\left({-1\over k^2}\right)^a
\left({-1\over l^2}\right)^b \left({-1\over (k-l)^2}\right)^c
\left({\omega\over v \cdot k+\omega}\right)^p
\left({\omega^{(')}\over v\cdot l +\omega^{(')}}\right)^q}.
\label{I_B}
\end{equation}
Those integrals $I$ depending on only one scale $\w$ have been
evaluated in~\cite{broadgro1}.
Only ${I'}(a,b,c,p,q)$ deserves some comments. For
$p>1$ and/or $q>1$,
${I'}(a,b,c,p,q)$ can be obtained
from ${I'}(a,b,c,1,1)$ by simply taking derivatives with respect
to $\omega$ and/or $\omega'$.
To evaluate
${I'}(a,b,c,1,1)$ one can make use of the following master equation:
\begin{eqnarray}
& &
\left[c-a+{\omega' \over \omega}(2a+c+p-D)\right] I'(a,b,c,p,q)
\label{master}
\\
&=&
\left[ c(1-{\omega' \over \omega}) C^+ (A^- - B^-) +aA^+(B^- -C^-)
+{\omega' \over \omega} p P^+ Q^-\right] I'(a,b,c,p,q),
\nn
\end{eqnarray}
where the action of $A^\pm$ is given by
$A^{\pm}I'(a,b,c,p,q)=I'(a\pm 1,b,c,p,q)$,
with similar definitions
of $B^\pm$, $C^\pm$, $P^\pm$ and $Q^\pm$.
We have deduced \eq.(master) by following the ``integration by parts''
procedure~\cite{master} also used in~\cite{broadgro1}.
By means of \eq.(master), one can reduce
any ${I'}(a,b,c,p,q)$ appearing in the calculation of $\gi$
to cases where at least one of the arguments
$a$, $b$, $c$, $p$, or $q$ vanishes. The cases $p=0$, $q=0$, $c=0$ can
be evaluated using the formul\ae\ given in \cite{broadgro1}. For the
cases $a=0$ and $b=0$ we make the change $k,l\rightarrow (l+k)$, and
use the results of \cite{broadgro1}
as well as
\begin{eqnarray}
\int d^Dk\;\left({-1\over k^2}\right)^a
\left({\omega \over \omega + v\cdot  k}
\right)^p
\left({\omega' \over \omega' + v\cdot  k}\right)^q
&=&i\pi^{D/2} {\Gamma(2a+p+q-D)\Gamma(D/2-a)\over \Gamma(a)
\Gamma(p+q)}\\ \times
\left({\omega \over \omega'}\right)^p (-2\omega')^{D-2a}\!\!\!& &
\!\!\! {_2 F_1} (p+q+2a-D,p,p+q,1-{\omega \over \omega' }),
\nn
\label{1loop us}
\end{eqnarray}
where $_2F_1(a,b,c,t)$ is the hypergeometric function.

Using a similar technique,
it is straightforward to obtain the lowest order contribution, $\go$.
Up to $O(\alpha_s)$, the full three-point function can be written as
\begin{equation}
\widetilde{\Gamma}=
{1\over\epsilon}\go_{\rm div}+\go_{\rm fin}+\epsilon\go_\ep
+{1\over\ep^2}\gi_{\rm div~II}+{1\over\epsilon}
\gi_{\rm div~I}+\gi_{\rm fin}
+\cdots ,
\label{Gfull}
\end{equation}
where the dependence in $\epsilon$ is given explicitely.
It is important to remember that the $O(\ep\,\alpha^0_s)$-terms
in \eq.(Gfull),
i.e $\go_\ep$,
cannot be neglected since they give a {\em finite} contribution of
order $\alpha_s$ to $\wt{\Gamma}$ through the external current
renormalization. The terms $\go_{\rm div}$,
$\gi_{\rm div~II}$ are polynomials in $\w$, $\w'$ and $r$ whereas the
other contributions in \eq.(Gfull)
contain logarithms and dilogarithms as well as
negative powers of $\w-\w'$.
By expanding these functions in powers of $\w'-\w$,
one can check that the limit $\w'\to\w$ of $\wt{\Gamma}$ exists.
Actually, $\widetilde{\Gamma}$ is analytic in $\w'-\w$ for $\w,~\w'<0$.
Note that the UV-divergent portion of $\gi$, i.e. $\gi_{\rm div~I}$
has got a non-polynomial structure in $\w$, $\w'$. As in ordinary QCD,
these non-localities disappear upon renormalization of the currents
$\wt J {}_5$ and $\wt V {}^\mu$. The corresponding renormalization
constants, $Z_{HL}$ and $Z_{HH}$ have been computed
by \cite{broadgro1,jimu,korrady} to be
\be
Z_{HL}=1+{\alpha_s\over 4\pi\epsilon}(-2),\quad
Z_{HH}=1+{\alpha_s\over 4\pi\epsilon}\left[{32\over 9}r+
{64\over 45}r^2+O(r^3) \right].
\label{Zs}
\ee
We have checked that the UV-divergent part of
the three-point function of the renormalized currents is a polynomial
in the external
variables $\w$, $\w'$, as it
should. This is a nontrivial check of our result. Hereafter,
$\wt\Gamma$ will stand for the {\em renormalized} three-point function.

As usual in the sum rule approach, one would like to express $\wt{\gam}$
as a dispersion integral, i.e.
\bea
\wt\Gamma (\omega, \omega',y) & = & -{1\over\pi^2}\int_0^\infty\!\!
ds\int_0^\infty\!\! ds'
{\rho_3(s,s',y)\over
(s-\omega)(s'-\omega')}+ \mbox{subtraction terms}
\label{Gamma spect repr}\\
\mbox{with\ }
\rho_3(\omega, \omega',r) & = &
-\im\left\{\im \Gamma(\w,s'+i0,r)\Bigr|_{\w=s+i0}
\right\} .
\label{Im Im}
\eea
Applying \eq.(Im Im) to our result for $\widetilde{\Gamma}$ we obtain
\begin{eqnarray}
-{1\over\pi^2}      \rho_3(s,s'\!,r)\!\!
&=&
\!\!
Q_0(s,s'\!,r){d^6\over ds^6}\delta(s-s')
\label{rho3}
\\
\!\!&+&\!\!
Q_1(s,s'\!,r){d^7\over ds^7}\!\log\vert s\!-\!s' \vert\!
+\!
Q_2(s,s'\!,r){d^7\over ds^7}[\;{\rm sgn}(s-s')\log\vert s\!-\!s'
\vert\;]\nn ,
\end{eqnarray}
where $Q_{0}$, $Q_{1}$, $Q_{2}$ are polynomials in $s$, $s'$ and $r$.
By substituting
the explicit result summarized in \eq.(rho3) in
the partial derivative of \eq.(Gamma spect repr)
with respect to $\w$ and $\w'$,
we have checked that one recovers the full expression for
$\partial^2\widetilde{\Gamma}(\w,\w',r)/\partial\w\partial\w'$
computed directly by differentiating our original result
for $\wt\Gamma (\w,\w',r)$. The differentiation is necessary
in order to eliminate the subtraction
terms in \eq.(Gamma spect repr).

Using the explicit form of \eq.(rho3) it is quite straightforward
to compute a quantity that plays a central role in the next section:
the Borel
transform of $\wt\Gamma $, ${\hat B}^{M_3}{\hat B}'^{M_3}\wt\Gamma$.
One just has to evaluate the integral
\begin{equation}
{\hat B}^{M_3}{\hat B}'^{M_3}\wt\Gamma=-{1\over\pi^2M_3^2}
\int_0^{\infty}\,ds\int_0^{\infty}\,ds'
\rho_3(s,s',y)\e^{-{(s+s')/ M_3}}, \label{rho3 int}
\end{equation}
where $M_3$ is the Borel parameter for the three-point function.
The final result is a very simple expression which
can be cast into the following form:
\begin{eqnarray}
{\hat B}^{M_3}{\hat B}'^{M_3}\wt\Gamma&=&
{2\over\pi^2M^2_3}\int_0^{\infty} ds\,s^2
{\rm e}{}^{-2s/M_3}
\left\{
{{3}\over {8}} +
    \ALPI\left[ {{17}\over {8}} -
      {{3\log (2s/\mu)}\over {4}} +
      {{\Z(2)}} \right]  \right.\nonumber
\\& &+    r
\left\{-{3\over4} +
      \ALPI\left[-{355\over72} +
         {{4\log 2}\over {3}}
         +{13\over6} \log (2s/\mu)-2 \Z(2) \right]  \right\}
\nonumber
\\
& &+\left.
  {r^2}
\left\{  {3\over8}+
      \ALPI  \left[   {11873\over3600}-
        {32\over15}\log 2 -{109\over60} \log (2s/\mu)
+
\Z(2)   \right]  \right\}
\right\}.
\label{B3}
\end{eqnarray}

\section{Evaluation of the Isgur-Wise function}\label{evaluation}

We are now in a position to derive a sum rule for the Isgur-Wise
function including the lowest order radiative
corrections. Sum rules for the meson-to-vacuum amplitudes
in the HQET were introduced by \cite{broadgro2,B3D} and applied
to the Isgur-Wise function by \cite{ball,neubert_physrev,rady}.
We will not go into details about the method of HQET sum rules,
for which we refer to \cite{B3D}, but simply review the main points.

One can saturate
eq.~(\ref{eq:correlationF}) with physical states of the HQET yielding
\bea
\wt{\gam}(v+v')_\mu \ = \ \wt{\gam}^{had}_\mu + \wt{\gam}^{cont}_\mu &
= & \frac{\langle\, 0 \, | \, \wt{J}_5 \, | \, \wt{P}'(v') \, \rangle
\langle \, \wt{P}'(v') \, | \, \wt{V}_\mu \, | \, \wt{P}(v) \,\rangle
\langle \, \wt{P}(v) \, | \, \wt{J}_5 \, | \, 0 \,\rangle}{4 (\dm -
\omega)(\dm-\omega')}\nn\\
& & {} + \mbox{contributions of higher states,}\label{eq:saturation}
\eea
where $\dm = m_M - m_Q\approx 0.5\gev$ is the difference between the
meson and the quark mass in the heavy quark limit (HQL). The matrix
elements in the numerator can be expressed in terms of the leptonic
decay constant in the HQL, $\wt{f}$,
\be
\langle\, 0 \, | \, \wt{J}_5 \, | \, \wt{P} \, \rangle = \wt{f},
\ee
and the universal Isgur-Wise function
\be
\langle \, \wt{P}' \, | \, \wt{V}_\mu \, | \, \wt{P} \,\rangle =
\xi_{IW}(y) (v+v')_\mu.
\ee
On the other hand, $\wt{\gam}$ can be calculated by means of the
operator product expansion (OPE) \`{a} la \cite{SVZ}, away from the
physical region, i.e.\ for $\omega,\,\omega' \ll 0$. Both expressions
are matched after application of the Borel transformation defined in
(\ref{rho3 int}) which suppresses the contributions of both higher
operators in the OPE and higher states in (\ref{eq:saturation}).

Due to the scale-dependence of $\wh{B}^{M_3}\,\wh{B}'^{M_3}\,\wt{\gam}$,
 we shall derive a sum rule for the renormalization group independent
(to two-loop accuracy) quantity
\be\label{eq:defgammahat}
\wh{B}^{M_3}\,\wh{B}'^{M_3}\,\wh{\gam} = \wh{B}^{M_3}\,\wh{B}'^{M_3}\,
\wt{\gam}(\mu)\alpha_s(\mu)^{-(2\ga_0^{HL} + \ga_0^{HH})/(2\beta_0)}\,
\left( 1 - \frac{\alpha_s(\mu)}{\pi}\left\{2 \Delta^{HL} +
\Delta^{HH}\right\} \right)
\ee
\be
\mbox{with\ }\Delta = \frac{\ga_0}{8\beta_0}\left( \frac{\ga_1}{\ga_0}
- \frac{\beta_1}{\beta_0} \right)\quad\mbox{and\ }\ga^{HH(HL)} =
\frac{\alpha_s}{4\pi}\left(\ga_0^{HH(HL)} + \frac{\alpha_s}{4\pi}
\ga_1^{HH(HL)} + \dots\right),
\ee
where $\ga^{HH}$ ($\ga^{HL}$) denotes the anomalous dimension of the
effective current with two heavy ( one heavy and one light) quarks
\cite{broadgro1,jimu,korrady}. $\beta_0$ and $\beta_1$
are the lowest order coefficients of the usual $\beta$-function of QCD.
Following the analysis of
\cite{B3D}, we choose $\mu = M_3$ which is the natural scale for the
Borel transformed expression.

As for the OPE, we take into account operators up to dimension 5 or,
to be specific, the quark condensate $\quark$ and the mixed condensate
$\gemischt$. The gluon condensate gives a small contribution and only at
$O(y-1)$ \cite{neubert_physrev}. Likewise, the contribution of the
four-quark condensate is expected to be small. Thus we write
\be\label{eq:OPE}
\wh{B}^{M_3}\,\wh{B}'^{M_3}\,\wh{\gam} = \wh{\gam}^{pert} +
\wh{\gam}^{\langle 3 \rangle} {\cal O}_3 +
\wh{\gam}^{\langle 5 \rangle} {\cal O}_5 + \dots
\ee
with the renormalization group invariant quark and mixed condensate
${\cal O}_3$ and ${\cal O}_5$, respectively, as defined in \cite{B3D}.
The perturbative contribution follows from (\ref{B3}):
\bea
\wh{\gam}^{pert} & = & \alpha_s(M_3)^{-(2\ga_0^{HL} + \ga_0^{HH}(y))/
(2\beta_0)} \frac{3}{\pi^2M_3^2 (1+y)^2} \intl_0^\infty\!\! ds\,s^2\,
e^{-2s/M_3}\nn\\
& & \left\{1+\frac{\alpha_s(M_3)}{\pi}\left[\left(\frac{\ga_0^{HL}}{2}
+ \frac{\ga_0^{HH}(y)}{4}\right) \ln \frac{2s}{M_3} +
\frac{4}{9}\pi^2 - 2 \Delta^{HL} - \Delta^{HH}(y)\right.\right.\nn\\
& & \left.\left. {} + \frac{17}{3} + (y-1) \underbrace{\left(
\frac{16}{9}\ln 2 - \frac{49}{54}\right)}_{\approx +0.3} + (y-1)^2
\underbrace{\left( {}-\frac{8}{15}\ln 2 + \frac{197}{600}\right)
}_{\approx -0.04}\right]\right\}.\label{eq:gammapert}
\eea
Here we have factored out an overall factor
$1/(y+1)^2$ which appears in the exact expression for the bare loop
diagram. Furthermore, we have summed up the coefficients in front of the
logarithm to get the anomalous dimension as specified by the
scale-dependence of $\wt{\gam}$. The numbers under the underbraces
show the good convergence of the series in $(y-1)$.

The lowest order expression for $\wh{\gam}^{\langle 3\rangle}$ is
simply given by $-1/(4M_3^2)$, whereas for the $O(\alpha_s)$
corrections we take the expression given by \cite{neubert_physrev}.
Radiative corrections to $\wh{\gam}^{\langle 5\rangle}$ have not been
calculated so far, so we use the lowest order expression
$\wh{\gam}^{\langle 5\rangle} = (2y+1)/(192M_3^2)$.

Using the usual argument of quark-hadron duality, the contribution of
higher states to (\ref{eq:saturation}) is modelled by the perturbative
contribution to the OPE (\ref{eq:OPE}) above certain threshold $\ds$,
 the so-called continuum threshold. Following the discussion of
\cite{blokshif}, we choose the threshold in such a way that
undesirable contributions of P-waves or even higher states do not mix
in. It was already noticed by \cite{ball,neubert_physrev} that the slope
 of the Isgur-Wise function is quite sensitive to the choice of the
continuum model and introduces a rather large uncertainty. This problem
gets resolved by choosing the continuum model according to
\cite{blokshif} as
\be\label{eq:continuummodel}
\wh{\gam}^{cont} = \intl_{\ds_3}^\infty \!\! ds \,
\times\mbox{\ Integrand of (\ref{eq:gammapert}).}
\ee
Taking all together we get the following sum rule for the
renormalization group invariant Isgur-Wise function:
\be\label{eq:SRIW}
\wh{\xi}_{IW}(y) = \frac{4 M_3^2\,(\wh{B}^{M_3}\,\wh{B}'^{M_3}\,
\wh{\gam}(y)-\wh{\gam}^{cont})}{\wh{B}^{M_2}\,\wh{SR}}
\ee
where $\wh{B}^{M_2}\,\wh{SR}$ denotes the corresponding sum rule for
the square of the leptonic decay constant, $\wh{f}^2$, and can be found
 in \cite{B3D}. Note that the exponential $\exp (-\dm/M_3)$, that
formally appears on the left hand side of (\ref{eq:SRIW}) due to the
Borel transformation, cancels against the
corresponding factor in $\wh{B}^{M_2}\,\wh{SR}$ and, hence
$\wh{\xi}_{IW}$ as calculated in (\ref{eq:SRIW}) becomes independent
of $\dm$. The sum rule (\ref{eq:SRIW}) automatically fulfills the
normalization condition $\left.\wh{\xi}_{IW}(y)\right|_{y=1} = 1$
for any value of
continuum threshold and the Borel parameter provided we take the same
value for $\ds_2 \equiv \ds_3 \equiv :\ds$ in both the numerator and
denominator and take $M_3 \equiv 2 M_2 \equiv :2 M$. This equivalence
will be used throughout the next section.

\section{Results and Discussion}\label{discussion}

We now turn to the numerical analysis of the sum rule
eq.~(\ref{eq:SRIW}). The numerical values of the condensates we use are
$\quark = (-0.24\gev)^3$ and $\gemischt = (0.8 \gev^2)\quark$ at
the scale of
$1\gev$. In Fig.~\ref{fig:parameters} we show $\wh{\xi}_{IW}$ for
different values of $\ds$ and $M$ as a function of $y$. For
phenomenological applications, such as the decay $B\to D^* e \nu$, it
is sufficient to know the Isgur-Wise function within the range
$1\leq y \leq 2$. In this region, the sensitivity of eq.~(\ref{eq:SRIW})
 to the continuum threshold and the Borel parameter is found to be quite
 small and to amount to at most 6\% for $y=2$. In the evaluation we have
taken $1.1\gev\leq \ds \leq 1.4\gev$ and $0.5\gev \leq M \leq 1.0 \gev$
as suggested by the analysis of the sum rule for $\wh{f}$ \cite{B3D}.
The resulting curves may be parametrized by a second order polynomial in
$(y-1)$ as
\be
\wh{\xi}_{IW}(y) = 1 - (0.54\pm 0.01) (y-1) + (0.17 \pm 0.01) (y-1)^2
\ee
where the errors reflect the uncertainty due to $M$ and $\ds$. The
result is mainly determined by the perturbative term which contributes
between 50\% and 85\% at $y=1$, depending strongly on $M$, and
increasing with $y$.

In Fig.~\ref{fig:radcorr} we depict the effect of neglecting radiative
corrections which is less than 10\% for $y\leq 2$. It is, however,
clearly seen that the inclusion of radiative corrections lowers the
values of $\wh{\xi}_{IW}$. This is not unexpected an effect, since the
coefficients of powers of $(y-1)$ in eq.~(\ref{eq:gammapert}) are quite
small and overcome by the $y$-dependent terms in front of
the logarithm. On the other hand, the large constant
term $\sim \alpha_s/\pi\,(17/3 + 4\pi^2/9)$, that was attributed to
Coulombic corrections in \cite{B3D}, is cancelled to a large extent by
the corresponding term in $\wh{B}^{M_2}\,\wh{SR}$. In other words:
the radiative corrections are mainly determined by the one-gluon
exchange between the heavy-quark lines, i.e.\ radiative corrections to
the weak vertex, and thus pure quantum effects (and not classical ones
such as Coulombic corrections).

In order to determine the slope parameter $\rho^2$ of the Isgur-Wise
function, $\rho^2 = {}-\left.(d\xi_{IW}(y))/(dy)\right|_{y=1}$, we scale
$\wh{\xi}_{IW}$ to a physically meaningful scale as
\be
\xi_{IW}(y,\bar m) = \left[ \al_s(\bar m)
\right]^{\ga_0^{HH}/(2\beta_0)}\left( 1 +
 \frac{\al_s(\bar m)}{\pi} \Delta^{HH} \right) \wh{\xi}_{IW}(y)\, .
\ee
To be specific, we choose $\bar m = m_B m_D/(m_B+m_D) \approx 1.4\gev$,
the harmonic mean of the masses of the B- and D-meson.

{}From the above formula we get $\rho^2 = (0.84\pm 0.02)$ where the error
stems from the uncertainty in the choice of parameters. This value is
nearly twice as big as the upper bound $\rho^2_{max} = 0.45$ given by
\cite{raftaron}. Disregarding all radiative corrections but the
overall leading-log scaling factors, we get $\rho^2 = (0.78\pm 0.02)$
which still is far away from $\rho^2_{max}$. We see no possibility to
get such a small value for the slope parameter, but rather agree with
the existing calculations and fits \cite{neubert_vcb} which yield
$\rho^2 \approx 1.0-1.5$. The resolution of this disagreement surely
requires further investigations.
\newpage
\vspace{1cm}

{\bf Acknowledgements}: We thank H.G.\ Dosch for some enlightening
discussions. P.G.\ acknowlegdes gratefully a grant from the Generalitat
 de Catalunya. E.B.\ acknowledges financial support of the CICYT.

\newpage

\section*{Figure Captions}

\bef[h]
\caption[]{$\wh{\xi}_{IW}$ as a function of $y$, eq.~(\ref{eq:SRIW}), for
$0.5\gev\leq M\leq 1.1\gev$, $1.1\gev\leq\ds\leq1.4\gev$. The spread of
the lines reflects the uncertainty due to
the choice of parameters.}\label{fig:parameters}
\enf

\bef[h]
\caption[]{Solid line: $\wh{\xi}_{IW}$ as a function of $y$,
eq.~(\ref{eq:SRIW}), at $M = 0.9\gev$, $\ds = 1.23\gev$. Dashed line:
$\wh{\xi}_{IW}$ without radiative corrections. The influence of
$O(\alpha_s)$ corrections causes a stronger falling-off of the form
factor mainly due to the corrections to the weak vertex.}
\label{fig:radcorr}
\enf

\end{document}